\documentclass[12pt]{article}
\usepackage[a4paper, margin=1in]{geometry}
\usepackage{times}
\usepackage{setspace}
\usepackage{bm}
\usepackage{titling}
\usepackage{abstract}
\usepackage{tikz}
\usepackage{hyperref}
\usepackage{graphicx}
\usepackage{amsmath, amssymb}
\usepackage{cite} 
\usepackage[numbers,sort&compress]{natbib}  

\title{Mean-field and Monte Carlo Analysis of Multi-Species Dynamics of agents}

\author{\parbox{\textwidth}{
  \centering Eduardo Velasco Stock, Roberto da Silva, Sebastian Gon\c{c}alves\\
  Instituto de F\'{i}sica, Universidade Federal do Rio Grande do Sul,\\ Caixa Postal 15051, 91501-970 Porto Alegre RS, Brazil\\
  \texttt{rdasilva@if.ufrgs.br} (Corresponding author)
}}

\date{\today}

\newcommand{\R}{\color{red}}

\begin{document}

\maketitle

\begin{abstract}
{We propose a mean-field (MF) approximation for the recurrence relation governing the dynamics of $m$ species of particles on a square lattice, and we simultaneously perform Monte Carlo (MC) simulations under identical initial conditions to emulate the intricate motion observed in environments such as subway corridors and scramble crossings in large cities. Each species moves according to transition probabilities influenced by its respective static floor field and the state of neighboring cells. To illustrate the methodology, we analyze statistical fluctuations in the spatial distribution for $m = 1$, $m = 2$, and $m = 4$ and for different regimes of average density and biased movement. A numerical comparison is conducted to determine the best agreement between the MC simulations and the MF approximation considering a renormalization exponent $\beta$ that optimizes the fit between methods. Finally, we report a phenomenon we term "Gaussian-to-Gaussian" behavior, in which an initially normal distribution of particles becomes distorted due to interactions among same and opposing species, passes through a transient regime, and eventually returns to a Gaussian-like profile in the steady state, after multiple rounds of motion under periodic boundary conditions.
}

\vspace{1em}
\end{abstract}

\section{Introduction}

Pedestrian dynamics is one example of complex systems of self-driven agents in the realm of macro-scale physics, which give rise to a variety of emerging phenomena. Due to its importance, a several theoretical and experimental studies have been done on crowd dynamics with a focus ranging from city planning~\cite{guo2023} and disaster prevention to the knowledge of pure intricate human behavior~\cite{Geoerg2022,Nicholls2024}.

Some places with crowd formation as subway corridors~\cite{ZHANG2022127656,HUANG2024103359}, nightclub dynamics~\cite{stock2023,Stock2024}, and scramble crossings~\cite{scramble2024} in large cities, are examples of these complex systems that exhibit collective behavior. Simply put, it features different groups of individuals aiming to reach various common destinations, for instance, the entrance/exit of a train station, reaching/leaving the bar to buy beverages, or crossing to get to another corner. Thus, the spatial and temporal context in question may involve a single group of agents moving towards a common goal, or multiple groups moving along confronting trajectories, which rapidly increases the complexity of the dynamics.

The problem of two groups of particles in counterflow has been extensively studied across various contexts, ranging from microscopic scales—such as the dynamics of charged colloids~\cite{Vissers-band-formation-2011}—to macroscopic phenomena, such as pedestrian flows in corridors~\cite{mestrado-eduardo-roberto2017,Zhang-intersecting-pedestrian-flows-experiment-2014}, and even more complex situations like evacuation~\cite{Helbing-simulating-dynamics-features-escape-panic-2000} scenarios. Notably, even in single-species models~\cite{majundar_evans2005}, intriguing phenomena such as condensate formation emerge, illustrating the richness of these systems.

An interesting approach to describe the hard-body dynamics of macroscopic systems is the lattice-gas modelling~\cite{katz1983}. In our recent work, we used the dynamics of a lattice gas-type interaction to describe the motion of pedestrians inside nightclubs~\cite{stock2023,Stock2024} and in a four-way crossing walk~\cite {scramble2024}, the so-called scramble crossing. In such works, we have shown that the intrinsic high correlation of this approach reflects the model's sensitivity to jamming and condensate formation.

In that sense, Monte Carlo (MC) simulations are a natural method for studying such models, as they allow for the definition and implementation of transition rules to simulate different evolutions from specified initial conditions. These systems can also be interpreted in a broader context as multi-agent systems. Additionally, there are approaches that model these dynamics through differential equations based on Newton's laws, within the so-called social force framework~\cite{helbing-1995,Helbing-simulating-dynamics-features-escape-panic-2000,oliveira2015}.

In that direction, the mean-field (MF) approach in particle dynamics has been extensively studied, particularly in the context of lattice gas models. These models were originally proposed to describe the behavior of specific types of fluids, such as solvents~\cite{KLEINTJENS1988}, electrochemical cells~\cite{Bernard2003}, and other systems that can be characterized by Hamiltonians derived from free energy potentials~\cite{dickman_lattice_gas1988}. In 2015, we introduced a related model~\cite{roberto2015} based on a two-species particle framework to describe pedestrian counterflow. Unlike classical exclusion principles, our model employed density-dependent exclusion rules, where the probability of occupation varied with the local density of particles. The system's evolution was governed by a set of coupled partial differential equations (PDEs). Other variations have been explored to model deterministic dynamics~\cite{Cividini_2013,APPERTROLLAND2014}, and further extensions of our framework have incorporated concepts from Fermi-Dirac statistics~\cite{stock2020}.

In this work, we propose a general framework for a MF approximation of the lattice gas dynamics of m-species of particles on a square lattice. We derive a generalized non-linear coupled PDE obtained by the MF approximation and compare it to MC simulations, by using a set of initial conditions for the cases $m=1$, $m=2$, and $m=4$, where each type of particle is guided by its own static floor field~\cite{schad2018} towards its target. We study the statistical properties of the distributions of particles of the species and investigate what are the circumstances and range of parameters that make the model a suitable and efficient option for MC simulations.

Our results show excellent agreement between the PDE predictions and MC simulations for low-density regimes. However, at higher densities, jamming effects reduce the accuracy of this agreement. We present a detailed numerical study to delineate the conditions under which our method remains valid compared to the computationally more expensive MC simulations.

In the next section, we describe the method for deriving the PDEs based on a model where particles exhibit preferential directed motion combined with random movements dependent on the occupation of neighboring sites. Through a mean-field approximation, we successfully derive a generalized PDE that describes scenarios involving multiple (m) species.

Subsequently, in the third section, we present our results comparing the spatial distribution of particles on a bidimensional lattice for different model parameters and initial particle concentrations. We study three different cases ---$m = 1$, $m = 2$, and $m = 4$---, analyzing the evolution of key distribution parameters over time under periodic boundary conditions. These parameters govern the interactions between particles across the various scenarios. Additional details ---including equation derivations, supplementary plots, and a discussion of numerical instabilities related to the PDEs--- are provided in the supplemental material accompanying this text.

\section{Methods}

Our model consists of a system of $n$ particles that move along an underlying square lattice of side $l$. Each particle in the system belong to one of $m$ species and can hop to one of its first neighboring cells at each time step according to a set of transition probabilities which depends on its species ``social fields'' (static floor field), denoted as $\mathbf{u}^{(k)}=\mathbf{u}^{(k)}(\textbf{r})$, where $k=1,\cdots,m$.
For instance, a given particle belonging to the species $q$ at cell $\textbf{r}=(i,j,t)$ hops to one of its neighboring cells $\textbf{r}(t+1)^{\prime}=(i^{\prime},j^{\prime},t+1)$ in the unit time interval $t\rightarrow t+1$ according to the transition probability density
\begin{equation}
Pr_{\textbf{r}\rightarrow \textbf{r}^{\prime}}^{(q)}=p+\alpha  \Delta \textbf{r}\bullet \textbf{u}^{(q)},
\label{prob_mov}
\end{equation}%
where $p$ is a constant probability to hop to one of the four neighboring cells, $\alpha$ is the coefficient that measure the bias towards the direction of its species static floor field, $\textbf{u}^{(q)}\equiv \mathbf{u}^{(q)}/||\mathbf{u}^{(q)}||$ is the normalized static floor field of species $q$, and $\Delta \textbf{r} \equiv \textbf{r}^{\prime}-\textbf{r}$ is the displacement vector of the calculated transition probability. As one may expect, normalization constraints reflect on the possible values of $p$, which in a Cartesian regular lattice is $0<p\leq 1/4$, whilst $\alpha $ satisfies the condition $0<\alpha \leq p$.

It is important to note that, in our model, each particle perceives only the static floor field specific to its species, evaluated at its current position. Consequently, the inner product defined in Eq.~\ref{prob_mov} quantifies the contribution of each possible transition probability to the particle’s movement toward its target.
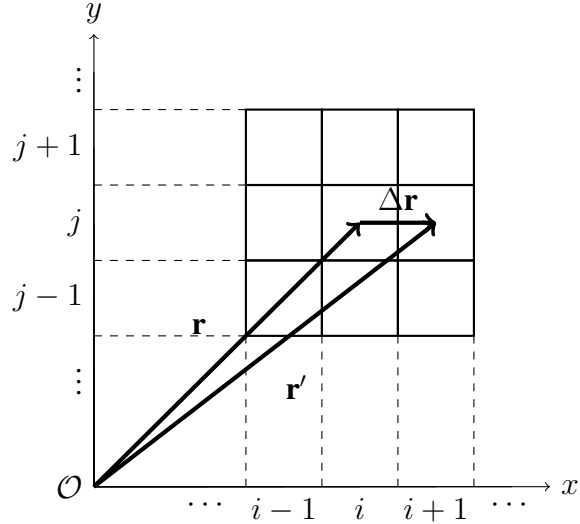
\begin{figure}
\centering
\begin{tikzpicture}
  
  \foreach \x in {0,1,2}
    \foreach \y in {0,1,2}
      \draw[thick] (\x+2,\y+2) rectangle (\x+3,\y+3);

  \foreach \x in {2,3,4,5}
    \draw[dashed] (\x,0) -- (\x,2);
  \foreach \y in {2,3,4,5}
    \draw[dashed] (0,\y) -- (2,\y);

  \coordinate (O) at (0,0);

  \coordinate (R) at (3.5, 3.5);
  
  \draw[->, ultra thick, black] (O) -- (R) node[midway, above left=3pt] {$\textbf{r}$};

  \coordinate (Rp) at (4.5, 3.5);

  \draw[->, ultra thick, black] (O) -- (Rp) node[midway, below right=3pt] {$\textbf{r}^{\prime}$};

  \draw[->, ultra thick, black] (R) -- (Rp) node[midway, above] {$\Delta\textbf{r}$};

  \draw[->] (0,0) -- (6,0) node[right] {$x$};
  \draw[->] (0,0) -- (0,6) node[above] {$y$};
  \draw[] (0,0) -- (0,0) node[left] {$\mathcal{O}$};
  \draw[] (0,0) -- (0,1.5) node[left] {$\vdots$};
  \draw[] (0,0) -- (0,2.5) node[left] {$j-1$};
  \draw[] (0,0) -- (0,3.5) node[left] {$j$};
  \draw[] (0,0) -- (0,4.5) node[left] {$j+1$};
  \draw[] (0,0) -- (0,5.5) node[left] {$\vdots$};
  \draw[] (0,0) -- (1.5,0) node[below] {$\cdots$};
  \draw[] (0,0) -- (2.5,0) node[below] {$i-1$};
  \draw[] (0,0) -- (3.5,0) node[below] {$i$};
  \draw[] (0,0) -- (4.5,0) node[below] {$i+1$};
  \draw[] (0,0) -- (5.5,0) node[below] {$\cdots$};
\end{tikzpicture}
\caption{Cartesian lattice depiction of a likely transition to happen of a particle at position $\textbf{r}$ to a neighbouring cell $\textbf{r}^{\prime}=\textbf{r}+\Delta \textbf{r}$.}
\label{fig1_displacement}
\end{figure}
As one might anticipate, Eq.~\ref{prob_mov} is the general form of the four possible transition probabilities each of which have a different corresponding displacement vector that can assume the values $ \Delta \textbf{r}=\textbf{e}_{x}$, $-\textbf{e}_{x}$, $\textbf{e}_{y}$, and $-\textbf{e}_{y}$, with $\textbf{e}_{x}$ and $\textbf{e}_{y}$ denoting the unit vectors of our reference frame as depicted in Figure~\ref{fig1_displacement}. Thus, to ensure the normalization constraint, the probability that a particle remains in its current cell is expressed as: 
\begin{equation}
Pr_{\textbf{r}\rightarrow \textbf{r}}^{(q)}=1-\sum\limits_{\left\langle\textbf{r}^{\prime} \right\rangle }Pr_{\textbf{r}\rightarrow \textbf{r}^{\prime}}^{(q)}=1-4p\text{,}
\label{prob_stay}
\end{equation}%
where the angle brackets $\left\langle \textbf{r}^{\prime} \right\rangle $ indicate that the summation is carried out solely over the nearest neighbours, with $i^{\prime} =i\pm 1$ and $j^{\prime} =j\pm 1$. 

It is important to note that these transition probabilities define the intended particle movement, conditional on the target cell being empty, as each cell can accommodate at most one particle. Thus, the probability represents what should occur, not what will necessarily occur.

\subsection{Lattice gas dynamics}

\label{Sec:lattice_gas}

We focus our study on an approximate regime of a lattice gas model. The key feature, as previously noted, is the exclusion principle, which prevents particles from overlapping or occupying the same cell, regardless of the specific rules governing the transition probabilities. This characteristic makes it particularly challenging to derive an approximate recurrence relation or equation of motion, as the occupation states of neighboring lattice cells are highly correlated. Nevertheless, we develop a recurrence relation for the cell occupation that remains consistent with the asynchronous updating scheme characteristic of lattice gas dynamics with nearest-neighbor interactions. Accordingly, we propose the following recurrence relation:
\begin{equation}
\begin{array}{lll}
\rho_{t+1}^{(q)}(\textbf{r})&=&\left(1-\rho_{t}(\textbf{r})\right)\sum\limits_{\langle \textbf{r}^{\prime}\rangle}\rho^{(q)}_{t}{(\textbf{r}^{\prime})}Pr_{\textbf{r}^{\prime}\rightarrow \textbf{r}}^{(q)}\\ &+&\rho_{t}^{(q)}(\textbf{r})\left(Pr_{\textbf{r}\rightarrow \textbf{r}}^{(q)}+\sum\limits_{\langle \textbf{r}^{\prime}\rangle}\rho_{t}(\textbf{r}^{\prime})Pr_{\textbf{r}\rightarrow \textbf{r}^{\prime}}^{(q)}\right),
\end{array}
\label{rec_1.1}
\end{equation}
where $\rho_{t}(\textbf{r})\equiv \sum\limits_{k=1}^{m}\rho_{t}^{(k)}(\textbf{r})$ is the density of particles of species $q$ at a given cell $\textbf{r}$ at time step $t$. The proposed equation decomposes the update process into two key contributions: (1) particle inflow into the cell and (2) particle persistence or outflow from the cell. In the lattice gas dynamics, the occupation state is binary, with $\rho_{t}^{(q)}(\textbf{r}) = 1$ if the cell is occupied and $\rho_{t}^{(q)}(\textbf{r}) = 0$ if it is vacant. However, the mean-field approximation accounts for the average behavior of $\rho_{t}(\textbf{r})$, which brings no actual information with respect to the actual number of particles of the system ($n$). Thus, we generalize the normalization constraint of distributions of particles of the recurrence relation such that the sum over all lattice sites scales as a power of $n$:
$$
\sum_{\textbf{r}} \rho_{t}(\textbf{r}) = n^{\beta},
$$  
where $\beta$ is an exponent that governs improvements in matching with MC simulations.

The first term on the right-hand side of Eq.~\ref{rec_1.1}, the factor $\left(1-\rho_{t}(\textbf{r})\right)$, reflects the hard-body exclusion principle, which dictates that a particle can only occupy an empty cell. This factor multiplies the sum of possible sources of inflow of particles of species $q$ from the neighboring cells of $\textbf{r}$, {\em i.e.} the notation $\langle \textbf{r}^{\prime}\rangle$. Each neighboring source of particles depends on the density of the species $q$ at the neighboring site $\rho^{(q)}_{t}(\textbf{r}^{\prime})$ and the transition probability $Pr_{\textbf{r}^{\prime}\rightarrow \textbf{r}}^{(q)}$, reflecting its likelihood to hop from $\textbf{r}^{\prime}$ to $\textbf{r}$. The second term on the right-hand side of Eq.~\ref{rec_1.1} describes the contribution of particles already present in $\textbf{r}$. The first factor, $\rho_{t}^{(q)}(\textbf{r})$, accounts for the actual occupation of it, whilst the factor inside the large brackets corresponds to a composition of two possibilities. The first possibility is the probability $Pr_{\textbf{r}\rightarrow \textbf{r}}^{(q)}$ of the particle at $\textbf{r}$ remaining in place, which contrasts with the second possibility that stands as the probability of the particle at $\textbf{r}$ trying to move to an occupied neighboring cell $\textbf{r}^{\prime}$.

With this framework, we explicitly separate the exclusion interaction, reflected in the terms $\left(1-\rho_{t}(\textbf{r})\right)$ and $\rho_{t}(\textbf{r})$, from the transition probabilities, which can incorporate additional interaction mechanisms such as long-range fields or cooperative dynamics. As an example, the Eq.~\ref{rec_1.1} can reproduce even deterministic models of particles, such as proposed in~\cite{Cividini_2013}, which could allow us to describe the model proposed by Cividini et al., for instance, by a proper choice of parameters such as fixing $m=2$ with one species moving eastbound and the other species moving northbound with transition probabilities equal to one towards their respective target directions.

We can rewrite Eq.~\ref{rec_1.1} as follows
\begin{equation}
\begin{array}{lll}
\rho_{t+1}^{(q)}(\textbf{r})&=&\sum\limits_{\langle \textbf{r}^{\prime}\rangle}\left[\left(\omega_{t}^{(q)}(\textbf{r},\textbf{r}^{\prime})-\omega_{t}^{(q)}(\textbf{r}^{\prime},\textbf{r})\right)\right.\\&-&\left.\left(\rho_{t}(\textbf{r})\omega_{t}^{(q)}(\textbf{r},\textbf{r}^{\prime})-\rho_{t}(\textbf{r}^{\prime})\omega_{t}^{(q)}(\textbf{r}^{\prime},\textbf{r})\right)\right] \\ &+& \rho_{t}^{(q)},
\end{array}
\label{rec_1.2}
\end{equation}
where $\omega_{t}^{(q)}(\textbf{r},\textbf{r}^{\prime}) \equiv \rho^{(q)}_{t}{(\textbf{r}^{\prime})}Pr_{\textbf{r}^{\prime}\rightarrow \textbf{r}}^{(q)}$ is the flow rate of particles of species $q$ entering the cell at $\textbf{r}$ from the cell $\textbf{r}^{\prime}$. Alternatively, we could write Eq.~\ref{rec_1.2} as
\begin{equation}
\rho_{t+1}^{(q)}(\textbf{r})= \Phi^{(q)}_t(\textbf{r})-\Psi^{(q)}_t(\textbf{r})+ \rho_{t}^{(q)},
\label{rec_1.3}
\end{equation}
where
\begin{equation}
\Phi^{(q)}_t(\textbf{r})\equiv \sum\limits_{\langle \textbf{r}^{\prime}\rangle}\left(\omega_{t}^{(q)}(\textbf{r},\textbf{r}^{\prime})-\omega_{t}^{(q)}(\textbf{r}^{\prime},\textbf{r})\right)
\end{equation}
is the net flow rate of particles of species $q$ into the cell at $\textbf{r}$ and
\begin{equation}
\Psi^{(q)}_t(\textbf{r})\equiv \sum\limits_{\langle \textbf{r}^{\prime}\rangle}\left(\rho_{t}(\textbf{r})\omega_{t}^{(q)}(\textbf{r},\textbf{r}^{\prime})-\rho_{t}(\textbf{r}^{\prime})\omega_{t}^{(q)}(\textbf{r}^{\prime},\textbf{r})\right)
\end{equation}
is the frustrated net flow rate of particles of species $q$ into the cell at $\textbf{r}$. The frustrated net flow rate measures the particle interaction, {\em i.e.}, measures the amount of $q$ particles unable access or leave the cell at $\textbf{r}$ due to the its occupation and of its neighbouring cells.  

\subsection{Mean-field regime}

\label{Sec:Mean-field}

When applying Eq.~\ref{prob_mov} into the lattice gas recurrence relation given by Eq.~\ref{rec_1.1}, we end up with the specific recurrence relation
\begin{equation}
\begin{array}{lll}
\rho_{t+1}^{(q)}(\textbf{r})&=&\left(1-\rho_{t}(\textbf{r})\right)\sum\limits_{\langle \textbf{r}^{\prime}\rangle}\rho^{(q)}_{t}{(\textbf{r}^{\prime})}\left(p-\alpha \left(\Delta \textbf{r}\bullet \textbf{u}^{(q)}(\textbf{r}^{\prime})\right)\right)\\ &+&\rho_{t}^{(q)}(\textbf{r})\left[\left(1-4p\right)+\sum\limits_{\langle \textbf{r}^{\prime}\rangle}\rho_{t}(\textbf{r}^{\prime})\left(p+\alpha \left(\Delta \textbf{r}\bullet \textbf{u}^{(q)}(\textbf{r})\right)\right)\right].
\end{array}
\label{rec_1.8}
\end{equation}

After some algebra (see \textcolor{blue}{Supplementary Material A}) we obtain the following partial differential equation
\begin{equation}
\begin{array}{lll}
\frac{\partial \rho^{(q)}}{\partial t}&=& c_1^{\prime} \nabla^2 \rho^{(q)}-c_2^{\prime}\nabla \bullet \left(\rho^{(q)}\textbf{u}^{(q)}\right)-c_3^{\prime}\sum\limits_{k\neq q}  \left(\rho^{(k)}\nabla^2 \rho^{(q)}- \rho^{(q)}\nabla^2 \rho^{(k)}\right)\\&+&c_4^{\prime}\sum\limits_{k=1}^{m}\left[\rho^{(k)}\nabla \bullet \left(\rho^{(q)}\textbf{u}^{(q)}\right)+\rho^{(q)}\nabla \rho^{(k)}\bullet \textbf{u}^{(q)}\right],
\end{array}
\label{edp_1.5}
\end{equation}
where $c_1^{\prime},c_3^{\prime}\propto \lim\limits_{\tau,\epsilon\rightarrow 0}\frac{p \epsilon^2}{\tau}$ and $c_2^{\prime},c_4^{\prime}\propto\lim\limits_{\tau,\epsilon\rightarrow 0}\frac{\alpha \epsilon}{\tau}$.

We notice that the constants $c_3^{\prime}$ and $c_4^{\prime}$ are species' coupling factors and the mathematical path we took from Eq.~\ref{rec_1.1} makes them not to depend on each possible pair of species. However, a more general approach would consider these constants as dependent on each possible coupling pair of species, which we will not going to explore in this work.

To study how good a fit is the mean-field approximation is to the particle dynamics, we investigate how the spatial distributions of both methods differ from each other for different sets of the model parameters such as the total number of particles ($n$), the total number of species ($m$), and the bias level of movement ($\alpha$).    

\section{Discussion}

As mean-field approaches approximate the microscopic behavior to an average behavior, their fidelity to the model tends to be limited to a certain range of parameters. This restriction is potentiated when considering cases of highly correlated systems of particles, such as the lattice gas dynamics we study in this work. In particular, we addressed how a system of an arbitrary number of species $m$, each carrying a different static and uniform social field $\textbf{u}$, influences emerging phenomena. As stated previously, the social field's direction is the only cause of difference between particles of different species, so that a system of $m>1$, but with $\textbf{u}^{(1)}=\textbf{u}^{(2)}=\cdots=\textbf{u}^{(m)}$ is not different from the case $m=1$. In that sense, the complexity of the model comes from different configurations of static floor fields.

Thus, to simplify things a little bit, we use a relatively simple rule to define social fields $\textbf{u}^{(q)}$, where each species has a guaranteed different field from each other in the two-dimensional Cartesian lattice:
\begin{equation}
    \textbf{u}^{(q)}=\left(\cos{\left(\frac{2\pi(q-1)}{m}\right)},\sin{\left(\frac{2\pi(q-1)}{m}\right)}\right).
    \label{static_floor_fields}
\end{equation}
By defining the static floor fields in Eq.~\ref{static_floor_fields} as such, we took the first species to have its preferential direction of movement parallel to the x-axis, whilst all other species' directions are defined in regularly divided angles according to the total number of particles ($m$).  

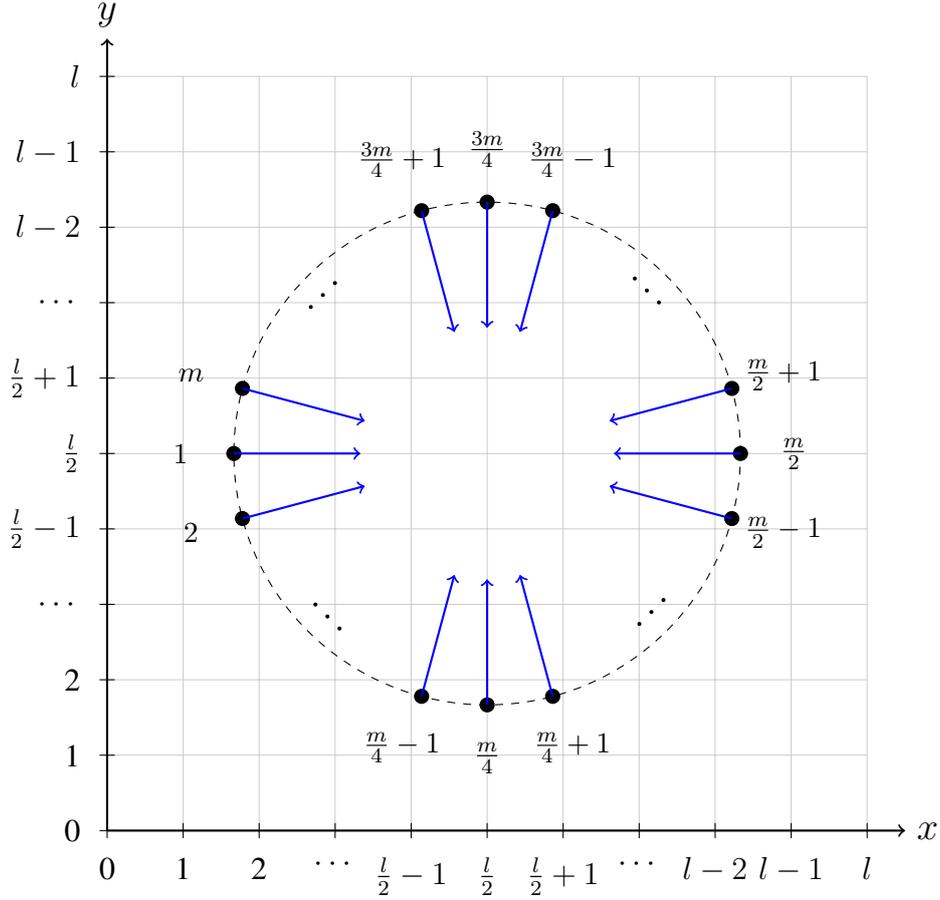
\begin{figure}
\centering
\begin{tikzpicture}

\def\l{10}
\def\m{24}
\def\R{\l/3}  
\def\cx{\l/2} 
\def\cy{\l/2} 

\draw[step=1cm, gray!40, very thin] (0,0) grid (\l,\l);
\draw[->, thick] (0,0) -- (\l+0.5,0) node[right] {\large$x$};
\draw[->, thick] (0,0) -- (0,\l+0.5) node[above] {\large$y$};

\draw (0,0.1) -- (0,-0.1) node[below=3pt] {0};
\draw (1,0.1) -- (1,-0.1) node[below=3pt] {1};
\draw (2,0.1) -- (2,-0.1) node[below=3pt] {2};
\draw (3,0.1) -- (3,-0.1) node[below=3pt] {$\cdots$};
\draw (4,0.1) -- (4,-0.1) node[below=3pt] {$\frac{l}{2}-1$};
\draw (5,0.1) -- (5,-0.1) node[below=3pt] {$\frac{l}{2}$};
\draw (6,0.1) -- (6,-0.1) node[below=3pt] {$\frac{l}{2}+1$};
\draw (7,0.1) -- (7,-0.1) node[below=3pt] {$\cdots$};
\draw (8,0.1) -- (8,-0.1) node[below=3pt] {$l-2$};
\draw (9,0.1) -- (9,-0.1) node[below=3pt] {$l-1$};
\draw (10,0.1) -- (10,-0.1) node[below=3pt] {$l$};

\draw (0.1,0) -- (-0.1,0) node[left=3pt] {0};
\draw (0.1,1) -- (-0.1,1) node[left=3pt] {1};
\draw (0.1,2) -- (-0.1,2) node[left=3pt] {2};
\draw (0.1,3) -- (-0.1,3) node[left=3pt] {$\cdots$};
\draw (0.1,4) -- (-0.1,4) node[left=3pt] {$\frac{l}{2}-1$};
\draw (0.1,5) -- (-0.1,5) node[left=3pt] {$\frac{l}{2}$};
\draw (0.1,6) -- (-0.1,6) node[left=3pt] {$\frac{l}{2}+1$};
\draw (0.1,7) -- (-0.1,7) node[left=3pt] {$\cdots$};
\draw (0.1,8) -- (-0.1,8) node[left=3pt] {$l-2$};
\draw (0.1,9) -- (-0.1,9) node[left=3pt] {$l-1$};
\draw (0.1,10) -- (-0.1,10) node[left=3pt] {$l$};

\draw[dashed] (\cx,\cy) circle (\R);

\foreach \q in {1,...,24} {
    \pgfmathsetmacro{\angle}{360*(\q-1)/\m}
    \pgfmathsetmacro{\x}{\cx - \R*cos(\angle)}
    \pgfmathsetmacro{\y}{\cy - \R*sin(\angle)}
    \pgfmathsetmacro{\vx}{\cx - 0.5*\R*cos(\angle)}
    \pgfmathsetmacro{\vy}{\cy - 0.5*\R*sin(\angle)}
    \ifnum\q=1
        \fill (\x,\y) circle[radius=0.1];
        \draw[->, thick, blue] (\x,\y) -- (\vx,\vy);
        \node[font=\small] at ({\x - 0.7*cos(\angle)}, {\y - 0.7*sin(\angle)}) {\small $1$};
    \fi
    \ifnum\q=2
        \fill (\x,\y) circle[radius=0.1];
        \draw[->, thick, blue] (\x,\y) -- (\vx,\vy);  
        \node[font=\small] at ({\x - 0.7*cos(\angle)}, {\y - 0.7*sin(\angle)}) {\small $2$};
    \fi
    \ifnum\q=6
        \fill (\x,\y) circle[radius=0.1];
        \draw[->, thick, blue] (\x,\y) -- (\vx,\vy);
        \node[font=\small] at ({\x - cos(\angle)}, {\y - 0.7*sin(\angle)}) {\small $\frac{m}{4}-1$};
    \fi
    \ifnum\q=7
        \fill (\x,\y) circle[radius=0.1];
        \draw[->, thick, blue] (\x,\y) -- (\vx,\vy);
        \node[font=\small] at ({\x - cos(\angle)}, {\y - 0.7*sin(\angle)}) {\small $\frac{m}{4}$};
    \fi
    \ifnum\q=8
        \fill (\x,\y) circle[radius=0.1];
        \draw[->, thick, blue] (\x,\y) -- (\vx,\vy);
        \node[font=\small] at ({\x - cos(\angle)}, {\y - 0.7*sin(\angle)}) {\small $\frac{m}{4}+1$};
    \fi
    \ifnum\q=12
        \fill (\x,\y) circle[radius=0.1];
        \draw[->, thick, blue] (\x,\y) -- (\vx,\vy);
        \node[font=\small] at ({\x - 0.7*cos(\angle)}, {\y - 0.7*sin(\angle)}) {\small $\frac{m}{2}-1$};
    \fi
    \ifnum\q=13
        \fill (\x,\y) circle[radius=0.1];
        \draw[->, thick, blue] (\x,\y) -- (\vx,\vy);
        \node[font=\small] at ({\x - 0.7*cos(\angle)}, {\y - 0.7*sin(\angle)}) {\small $\frac{m}{2}$};
    \fi
    \ifnum\q=14
        \fill (\x,\y) circle[radius=0.1];
        \draw[->, thick, blue] (\x,\y) -- (\vx,\vy);
        \node[font=\small] at ({\x - 0.7*cos(\angle)}, {\y - 0.7*sin(\angle)}) {\small $\frac{m}{2}+1$};
    \fi
    \ifnum\q=18
        \fill (\x,\y) circle[radius=0.1];
        \draw[->, thick, blue] (\x,\y) -- (\vx,\vy);
        \node[font=\small] at ({\x - cos(\angle)}, {\y - 0.7*sin(\angle)}) {\small $\frac{3m}{4}-1$};
    \fi
    \ifnum\q=19
        \fill (\x,\y) circle[radius=0.1];
        \draw[->, thick, blue] (\x,\y) -- (\vx,\vy);
        \node[font=\small] at ({\x - cos(\angle)}, {\y - 0.7*sin(\angle)}) {\small $\frac{3m}{4}$};
    \fi
    \ifnum\q=20
        \fill (\x,\y) circle[radius=0.1];
        \draw[->, thick, blue] (\x,\y) -- (\vx,\vy);
        \node[font=\small] at ({\x - cos(\angle)}, {\y - 0.7*sin(\angle)}) {\small $\frac{3m}{4}+1$};
    \fi
    \ifnum\q=24
        \fill (\x,\y) circle[radius=0.1];
        \draw[->, thick, blue] (\x,\y) -- (\vx,\vy);
        \node[font=\small] at ({\x - 0.7*cos(\angle)}, {\y - 0.7*sin(\angle)}) {\small $m$};
    \fi
}

\foreach \angle in {45,135,225,315} {
    \pgfmathsetmacro{\tx}{\cx - 0.9*\R*cos(\angle)}
    \pgfmathsetmacro{\ty}{\cy - 0.9*\R*sin(\angle)}
    \pgfmathsetmacro{\rot}{\angle + 90}
    \node[rotate=\rot] at (\tx,\ty) {\large$\cdots$};
}
\end{tikzpicture}
\caption{Schematic lattice (not to scale) illustrating a system with $m$ species in a situation involving confrontation. Each point marks the mean $\left(\mu_x^{(q)},\mu_y^{(q)}\right)$ of a species' initial distribution, with arrows indicating the direction of its static floor field. Angular separations are illustrative; actual values depend on $m$. The figure conveys the structure of initial conditions and static fields.}
\label{fig9_placement}
\end{figure}

As initial conditions, we considered each species concentrated in separate regions of the lattice as initial "wave packages" to describe what would be a real physical confrontation scenario, such as found in pedestrian crossings~\cite{scramble2024,Cividini_2013}. More specifically, we considered each species' initial distribution as a bivariate uncorrelated normal distribution with mean values defined in a circular fashion analogous to the static floor fields previously defined with a light difference. To observe confrontation between species, we defined the initial distributions of each as $\mu_x^{(q)}=\frac{l}{2}\left[1-\frac{1}{2}\cos{\left(\frac{2\pi(q-1)}{m}\right)}\right]$ and $\mu_y^{(q)}=\frac{l}{2}\left[1-\frac{1}{2}\sin{\left(\frac{2\pi(q-1)}{m}\right)}\right]$. That particular choice guarantees that agents will have to cross each other at the lattice center, simulating scenarios where species have well-defined positions and confrontation between groups of agents with different objectives is bound to occur as depicted in Fig.~\ref{fig9_placement}.

For all species, we define the standard deviation of the initial distributions to be $\sigma^{(q)}_x=\sigma^{(q)}_y=l/32$. The reason behind this specific choice for the standard deviation and not a smaller value is the constraint of the exclusion feature of the lattice model. For instance, a Dirac's delta function as initial distribution, where $\sigma_x^{(q)}=\sigma_y^{(q)}\rightarrow0$, can only be implemented on MC simulations if we put only one particle in the species mean initial position and we would have to make the system large enough so to have an approximate distribution of choice, which would be impractical computationally. In that sense, to study systems with more than one particle per species, we have to choose initial distributions not so localized.

Specifically, we have examined the cases of \( m = 1 \), \( m = 2 \), and \( m = 4 \), which represent simpler scenarios where the species' static floor fields align with the \( x \)- and \( y \)-axes. This alignment with the Cartesian axes provides a symmetric structure that facilitates a clearer comparison between the two methods via the marginalized distribution relative to the species' static floor field. For instance, if a species has a preferred direction \( \mathbf{u} = \pm \mathbf{e}_{x} \), we compute the marginalized density \( \rho(x) \) by summing \( \rho(x,y) \) over all \( y \)-values. This reduction to a one-dimensional distribution simplifies the analysis. 

In all cases, we studied a system of $l=128$ and $p=1/4$ with periodic boundary conditions on both directions (toroid). The latter reflects a scenario of high noise dynamics (see Eq.~\ref{prob_stay}) where particles only stay still if the target cell is occupied. For the MC simulations of the particle dynamics, we used a Box-Muller algorithm~\cite{Press1996} to generate two independent and normally distributed pseudo-random variables ($X$ and $Y$) to define each particle's position according to their species. Throughout this work, variables originating from MC simulations were obtained by averaging the time series of each run at each timestep over $n_{run}=10^6/n$ samples generated with same parameter, but different seeds for the pseudo-random number generators. 

\subsection{$m=1$ case}
\begin{figure}
\begin{center}
\includegraphics[trim={0cm 0cm 0cm 0cm},clip,width=1.0\textwidth]{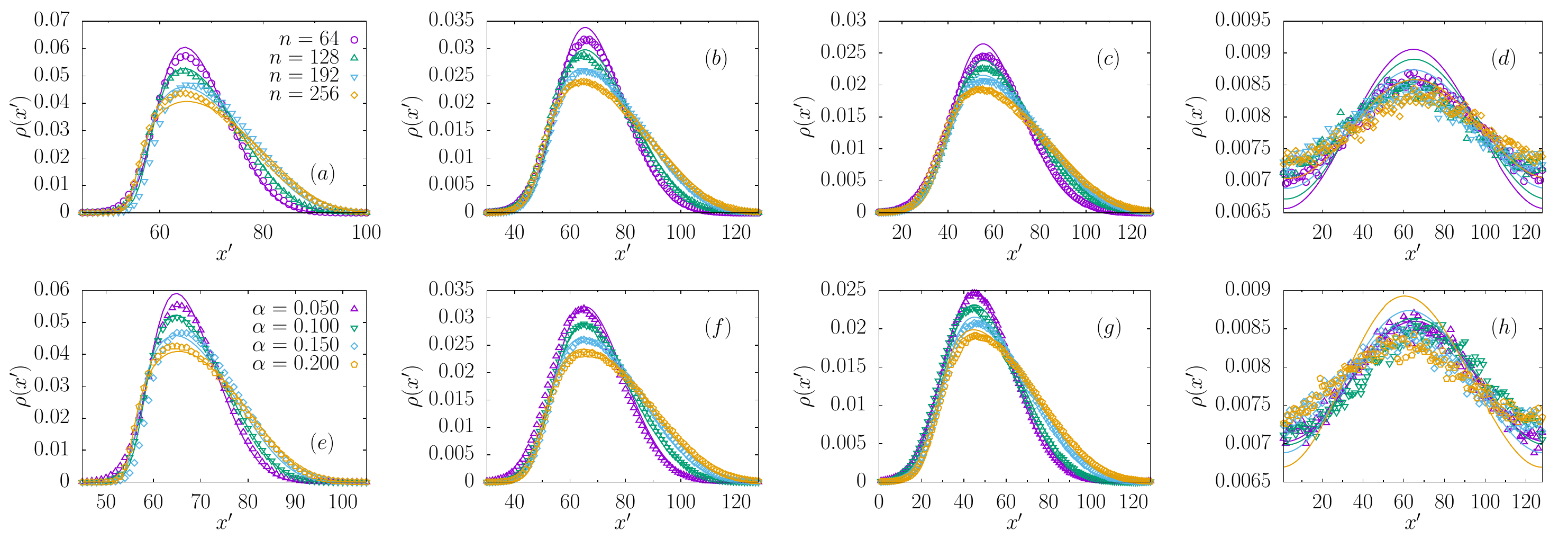}
\end{center}
\caption{Evolution of normalized marginal distributions of $x$ of MC simulations (circles) and MF (lines) for different time steps (a and e) $t=50$, (b and f) $t=250$, (c and g) $t=450$, and (d and h) $t=4950$ considering one-species ($m=1$). Plots labeled from (a) to (d) illustrate the influence of different values of $n$ for a fixed $\alpha = 0.15$, whereas plots from (e) to (h) show the effect of varying $\alpha$ for a fixed $n = 192$.}
\label{fig:m1}
\end{figure}

The simplest scenario we examined involves only a single species. As we defined previously by Eq.~\ref{static_floor_fields}, a single species tends to move according to the static floor field $\textbf{u}=\textbf{e}_x$, with its initial distribution having $\mu_x^{(1)}=l/4$ and $\mu_y^{(1)}=l/2$ with $\sigma^{(q)}_x=\sigma^{(q)}_y=l/32$. In these simulations, we used $\beta=1$.

In Fig.~\ref{fig:m1}, we present plots of the marginal distribution of particles along $x$, where circles represent the Monte Carlo (MC) simulation results and solid lines correspond to predictions from the recurrence relation (mean-field approximation). We examine these distributions for various particle counts, ranging from $n = 64$ to $n = 448$ in increments of $\Delta n = 64$ and shown at four distinct time steps: (a) $t = 50$, (b) $t = 250$, (c) $t = 450$, and (d) $t = 4850$. For this study, we used $\alpha=0.15$, which corresponds to agents with mid range level of impetus as we already shown in~\cite{Stock2024,scramble2024}.

As a general remark, we notice a qualitatively good agreement between the two methods and that the distributions in more densely populated systems shows more positive skewness of the probability density function (PDF) in comparison to systems with fewer particles, revealing the exclusion effects. These exclusion effects are a result of agents leading the pack acting as obstacles for the particles in the bulk. As time passes, we notice that the distributions present increasing dispersion with the ``front tail'' maintaining synchronicity for different values of $n$. It is important to disclosure that in the study that produced Fig.~\ref{fig:m1}, the recurrence relation exhibited numerical instability for $n \geq 512$. In some cases, certain values in the distribution reach magnitudes on the order of $10^{20}$, providing a quantitative indication that there exists a certain range of the parameters of the model beyond which the approximate method becomes unreliable and serves as justification for the assumption that a tuning parameter, $\beta$, might be necessary to also preserve the mean-field stability. The reader is advised to check the supplementary material to check the form of marginal distributions of particles right before numerical collapse.

Another key parameter we investigated is the bias movement level, $\alpha$, which quantifies the particles' tendency to move in their species' preferred direction. For this analysis, we fixed $n=192$, corresponding to a generally low-density scenario as shown, while ensuring sufficient particle interactions to produce the positive skewness observed in the distribution.

In the lower panels of Fig.~\ref{fig:m1} (plots e to h), we present the marginal distributions at time steps (e) $t=50$, (f) $t=250$, (g) $t=450$, and (h) $t=4950$ for $\alpha = 0.050, 0.100, 0.150,$ and $0.200$. As expected, distributions with higher $\alpha$ values exhibit faster drift velocities in the $+x$ direction. Additionally, increasing $\alpha$ amplifies the skewness, similar to the effect observed with higher $n$. Notably, at longer simulation times, larger $\alpha$ values lead to greater discrepancies between the two methods, as seen in Fig.~\ref{fig:m1} (h).
\begin{figure}
\begin{center}
\includegraphics[width=1.0\textwidth]{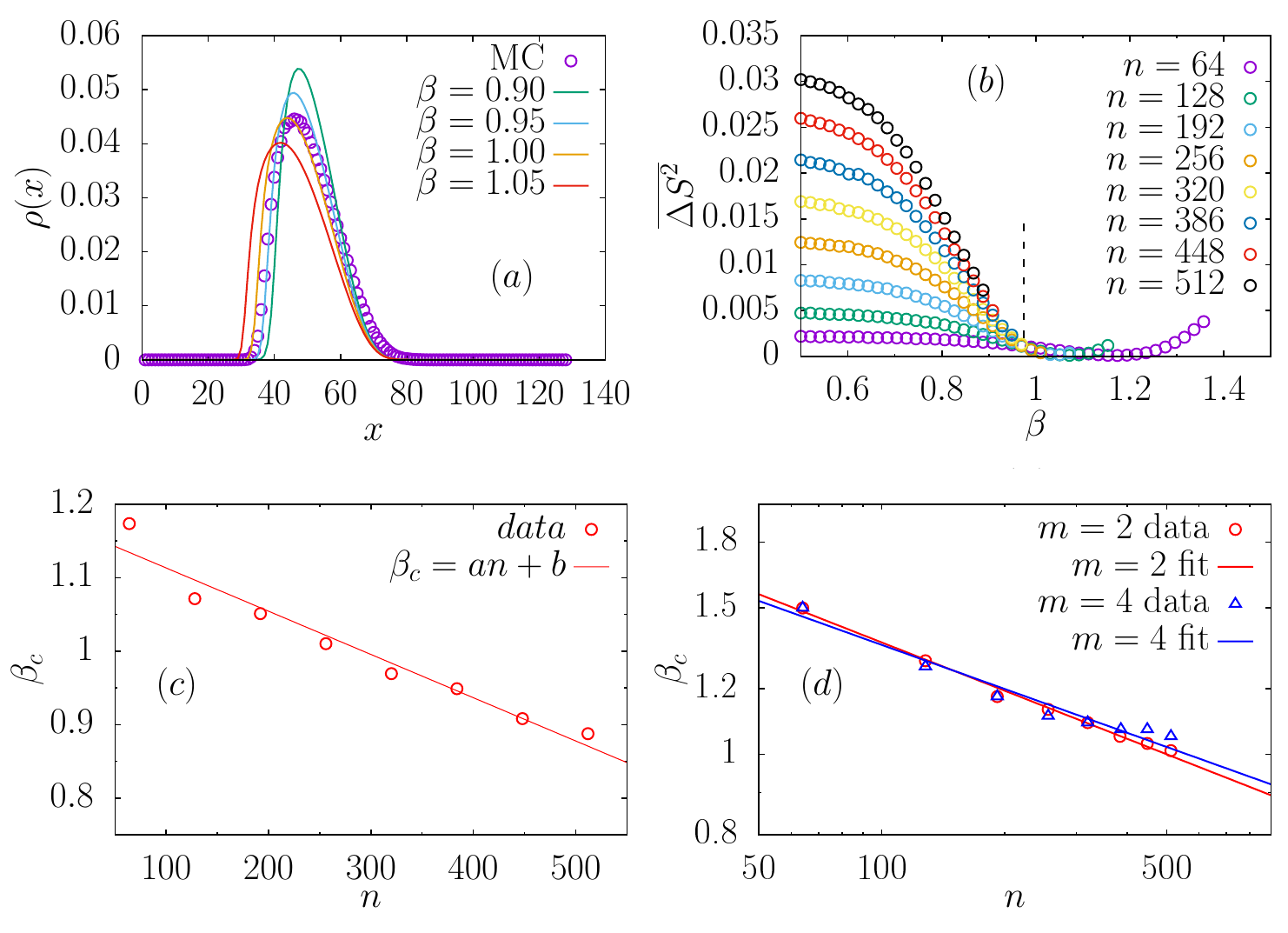}
\end{center}
\caption{(a) Influence of $\beta$ on the distribution profile for a system with parameters $m = 1$, $n = 128$, and $\alpha = 0.249$. (b) Variation of the average squared deviation $\overline{\Delta S^2}$ as a function of $\beta$, for a single-species system ($m = 1$) with $\alpha = 0.15$. (c) Completing the analysis for $m = 1$, we show the linear dependence of the optimal value $\beta_c$ on the number of agents. The data is fitted with a linear function $ax + b$, yielding $a = -0.000588 \pm 5.3 \times 10^{-5}$ and $b = 1.17 \pm 0.017$. (d) By extending the same calculations to systems with multiple species ($m = 2$ and $m = 4$), we observe that the results are well described by power function fits for $\beta_c$, using the form $ax^{-b}$. The fitted parameters are: $m = 2$: $a = 3.3 \pm 0.11$, $b = 0.193 \pm 0.0065$; $m = 4$: $a = 3.04 \pm 0.23$, $b = 0.176 \pm 0.014$.}
\label{fig:beta_study}
\end{figure}
Given the phenomena observed in Fig.~\ref{fig:m1}, we notice that systems with a greater total number of particles ($n$) tend to show less agreement, even if qualitatively their shapes follow the same pattern. However, this disparity is not observed when changing the level of movement bias ($\alpha$) shown in Fig.~\ref{fig:m1}. With this in mind, we show, in Fig.~\ref{fig:beta_study}, how the exponent of normalization, $\beta$, influences the shape of the marginal distribution of $x$. 

Fig.~\ref{fig:beta_study} (a) illustrates the influence of $\beta$ on the distribution profile for a system with parameters $m = 1$, $n = 128$, and $\alpha = 0.249$. As discussed in the previous section, the normalization constraint in the recurrence relation can be interpreted through an ansatz in which the number of particles is raised to a power $\beta$, with $\beta$ expected to be close to 1, if not exactly 1. As such, this parameter serves to tune the agreement 
between methods at the same time, it allows for systems with number of particles expected to show numerical collapse to evolve in a well-behaved manner.

To better measure the agreement between the two methods and the range of parameters where the mean-field shows numerical stability, we compared both methods using the time average of the squared difference of the spatial entropies of each method, {\em i.e.},
\begin{equation}
\overline{{\Delta S}^2}\equiv \frac{1}{t_{\max}}\sum_{t=1}^{t_{\max}}\left(S_{{t}_{\mbox{MC}}}-S_{{t}_{\mbox{MF}}}\right)^2,
\end{equation}
where $S_t=\sum\limits_{\textbf{r}}\rho_t(\textbf{r})\ln\rho_t(\textbf{r})$.

In Fig.~\ref{fig:beta_study} (b), we show $\overline{{\Delta S}^2}$ as a function of the exponent $\beta$ for different values of $n$ and for a fixed simulation time of $t_{\max}=10^4$ time steps. We studied the range of $\beta \in [0.5:1.5]$. In this range, we observed numerical instability of the recurrence relation in all curves shown for values of $\beta$ greater than certain values, which are reflected by the discontinuity of the curves. More specifically, greater $n$ shows lower $\beta$'s where the numerical collapse occurs. However, within the range of numerical stability, we observe that, for $n\le 256$, methods show an optimal agreement at $\beta\equiv \beta_c > 1$, shown by the local minima which become closer to 1 as $n$ increases. For $n\ge 256$, we notice that the numerical instability of the MF makes the lowest stable $\beta$ assume the stance of optimal value, {\em i.e.} it becomes the value that minimizes $\overline{{\Delta S}^2}$.

Fig.~\ref{fig:beta_study} (c) completes the analysis for $m = 1$, by showing the linear dependence of the optimal value $\beta_c$ on the number of agents. The data is fitted with a linear function $ax + b$, yielding $a = -0.000588 \pm 5.3 \times 10^{-5}$ and $b = 1.17 \pm 0.017$. Such behavior will be different of upper values of $m$ as it is previously presented in Fig.~\ref{fig:beta_study} (d). However, we must first study the particle distribution for these cases under varying parameters, which will be performed in the next sections.  

\subsection{$m=2$ case}

The case $m=2$ is the simplest case within our proposed framework of initial conditions (see Fig.~\ref{fig9_placement}) where two species of particles move in counterflow, which is a scenario vastly studied in the literature, once it can describe a two species systems with opposing responses to an external field such as oppositely charged coloids~\cite{Vissers-band-formation-2011,roberto2015} (external electrical field) or chromatographic columns~\cite{Gibbings-molecular-dynamic-theory-of-chromatography-1955,Roberto-combinatorial-method-particle-retention-time-chromatography-2012,mestrado-eduardo-roberto2017,Stock-JSTAT-2019,nossoPRE2019} (gravitational field), to cite a few. For this case, initial distributions assume $\mu_x^{(1)}=l/4$ and $\mu_y^{(1)}=l/2$, $\mu_x^{(2)}=3l/4$ and $\mu_y^{(2)}=l/2$ as we defined in the previous section. Static floor fields are then $\textbf{u}^{(1)}=\textbf{e}_x$ and $\textbf{u}^{(2)}=-\textbf{e}_x$ as stated by Eq.~\ref{static_floor_fields}, which still carries a rather trivial symmetry in the $y$-direction (diffusive profile) making observations of the marginal distribution on the $x$ coordinate rather more interesting because is where the confrontation occurs.
\begin{figure}
\begin{center}
\includegraphics[width=1.0\textwidth]{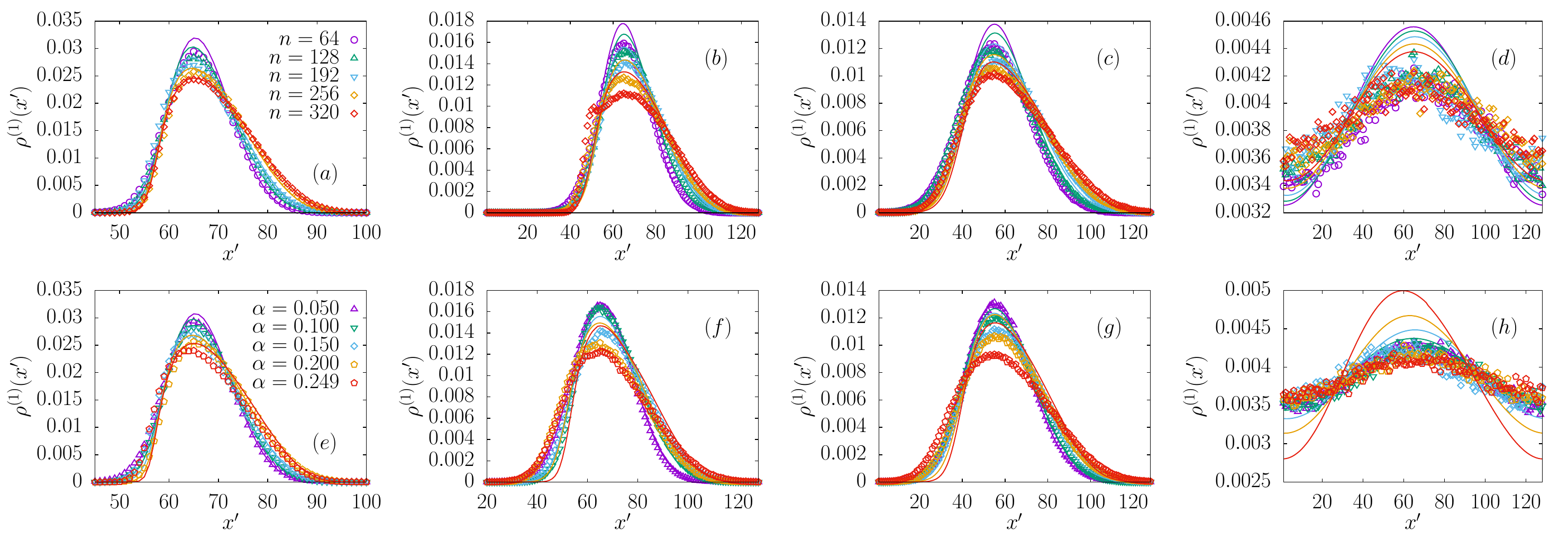}
\end{center}
\caption{Evolution of normalized marginal distributions of $x$ of the MC simulations (circles) and MF (lines) for different time steps (a and e) $t=50$, (b and f) $t=250$, (c and g) $t=450$, and (d and h) $t=4950$. Plots (a) to (d) show the influence of different values of $n$ for $\alpha=0.15$, whilst plots (e) to (h) show different $\alpha$'s for a fixed $n=192$.} \label{fig:m2}
\end{figure}

In Fig.~\ref{fig:m2}, we show the marginal distributions of $x$ of the MC simulations with circles and lines for the MF approach. We studied two values of the total number of particles shown simultaneously: $n=128$ shown in purple and $n=256$ shown in green for the time steps (a) $t=50$, (b) $t=250$, (c) $t=450$, and (d) $t=4850$, where both species appear as different peaks with same color for each case. In this first case, we used $\alpha=0.15$ and $\beta=1.0$, and we observe that, similarly to the $m=1$ case, a greater total number of particles makes the distributions more asymmetric due to the exclusion interaction of particles of the same species. We do not observe, however, a clear influence of the opposing species on the shape of the distribution. We can interpret this ``lack'' of interaction deformity occurring because of the exclusion effects of the nearest neighbor interaction, suggesting that a sort of shield is formed by the same-species particles in a low-density regime. This warrants further investigation in future work. 

While both methods show reasonably good agreement in Fig.~\ref{fig:m2}, we find that the agreement can be further improved by optimizing $\beta$. Revisiting Fig.~\ref{fig:beta_study} (d) and extending our earlier analysis for $m=1$ to $m=2$, we observe that the critical value $\beta_{c}$ minimizing $\overline{{\Delta S}^2}$ follows a power-law dependence, $ax^{-b}$. For $m=2$, the fit yields $a = 3.3 \pm 0.11$ and $b = 0.193 \pm 0.0065$, contrasting with the linear behavior found for $m=1$. This indicates that increasing the model complexity (from $m=1$ to $m=2$) modifies the parameter dependence. However, no significant changes are observed beyond $m=2$ up to $m=4$, as seen in Fig.~\ref{fig:beta_study} (d)—a point we will address in detail later. Importantly, we note that lower-density cases do not provide the best agreement between the two methods, revealing non-trivial model-specific characteristics that lack complete theoretical understanding.

\subsection{$m=4$ case}
\begin{figure}
\begin{center}
\includegraphics[width=1.0\textwidth]{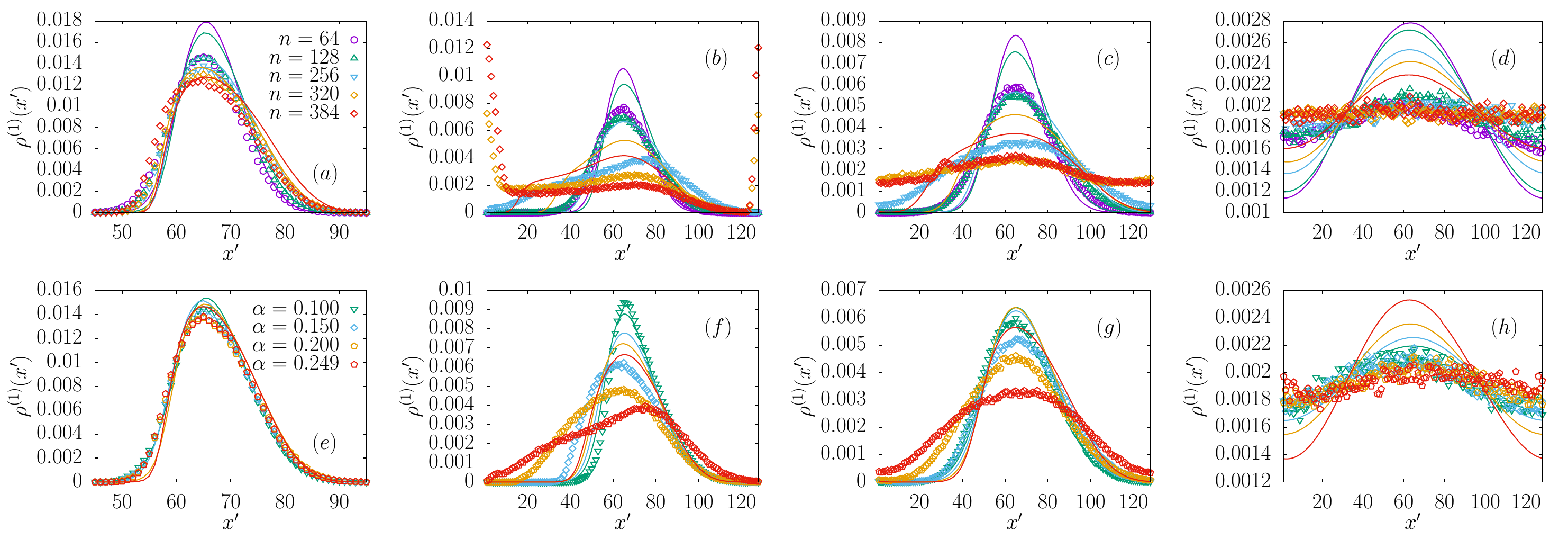}
\end{center}
\caption{Evolution of normalized marginal distributions of $x$ of the MC simulations (circles) and MF (lines) for different time steps (a and e) $t=50$, (b and f) $t=250$, (c and g) $t=450$, and (d and h) $t=4950$. Plots (a) to (d) show the influence of different values of $n$ for $\alpha=0.249$, whilst plots (e) to (h) show different $\alpha$'s for a fixed $n=256$.} \label{fig_m4}
\end{figure}

The last case we discuss in this work is the case with four species moving along the $x-$ and $y-$directions. More precisely, their initial distributions have mean value: $\mu_x^{(1)}=l/4$ and $\mu_y^{(1)}=l/2$, $\mu_x^{(2)}=3l/4$ and $\mu_y^{(2)}=l/2$, $\mu_x^{(3)}=l/2$ and $\mu_y^{(3)}=l/4$, and $\mu_x^{(4)}=l/2$ and $\mu_y^{(4)}=3l/2$ as we expected to be by our previous definition. Similarly, the static floor fields assume the form $\textbf{u}^{(1)}=\textbf{e}_x$, $\textbf{u}^{(2)}=\textbf{e}_y$, $\textbf{u}^{(3)}=-\textbf{e}_x$, and $\textbf{u}^{(4)}=-\textbf{e}_y$ (see Eq.~\ref{static_floor_fields}). Similarly to the two prior cases, having an initial condition with radial symmetry about the lattice center, also makes the four species present a symmetric ``cross section'' of the distribution, {\em i.e.} the distribution on the direction perpendicular to $\textbf{u}^{(q)}$ shows no asymmetry. Thus, studying the marginal distribution of $x$ of the species $q=1$ or $q=3$ would be no different from studying the marginal distribution of $y$ of the species $q=2$ and $q=4$ given the system's initial arrangement and static floor fields (see Fig.~\ref{fig9_placement}).

With this in mind, Figure~\ref{fig_m4} displays the particle distribution $\rho^{(1)}(x)$ for different values of $n$ using $\alpha=0.249$ at four time points: (a) $t=50$, (b) $t=250$, (c) $t=450$, and (d) $t=4850$. Our analysis reveals two key observations:

(1) First, the discrepancy between methods becomes more pronounced for larger $n$ values in this regime of highly driven agents. Despite this disagreement, the MC simulations demonstrate the formation of transient condensates for $n=320$, as evidenced by the emerging peak at $t=250$.

(2) Second, the mean-field solution exhibits anomalous behavior at $t=250$, particularly in the left tail of $\rho^{(1)}(x)$, which appears bent near $x\approx100$. Notably, however, both methods show qualitatively good agreement in the front tail of $\rho^{(1)}(x)$, suggesting synchronized drifting profiles in the steady state.

Building on our analysis of the $m=4$ case, Figure~\ref{fig_m4} presents the evolution of $\rho^{(1)}(x)$ distributions for various $\alpha$ values in a system with $n=256$ particles. We also examine the same four time points: (e) $t=50$, (f) $t=250$, (g) $t=450$, and (h) $t=4950$. The results reveal that while the wave fronts remain synchronized between MC simulations and mean-field theory, the MC distributions exhibit significantly greater dispersion. Returning to Fig.~\ref{fig:beta_study}(d), we apply the same minimization procedure for $\overline{{\Delta S}^2}$ used for the m=1 and m=2 cases. For m=4, we obtain an excellent power-law fit identical in form to the m=2 case, with fitted parameters a = 3.04 ± 0.23 and b = 0.176 ± 0.014.

It is important to note that, despite some differences in the evolution of particle profiles, a qualitatively similar behavior is observed in both cases when the systems are initialized with a normal distribution of particles in the environment. The interactions among particles—especially the interaction with counterflowing particles—tend to deform the initial distribution shape. Over time, however, after several rounds of confrontation and interaction, the particles begin to reorganize themselves, ultimately restoring a Gaussian-like behavior.

\begin{figure}
\begin{center}
\includegraphics[width=0.95\textwidth]{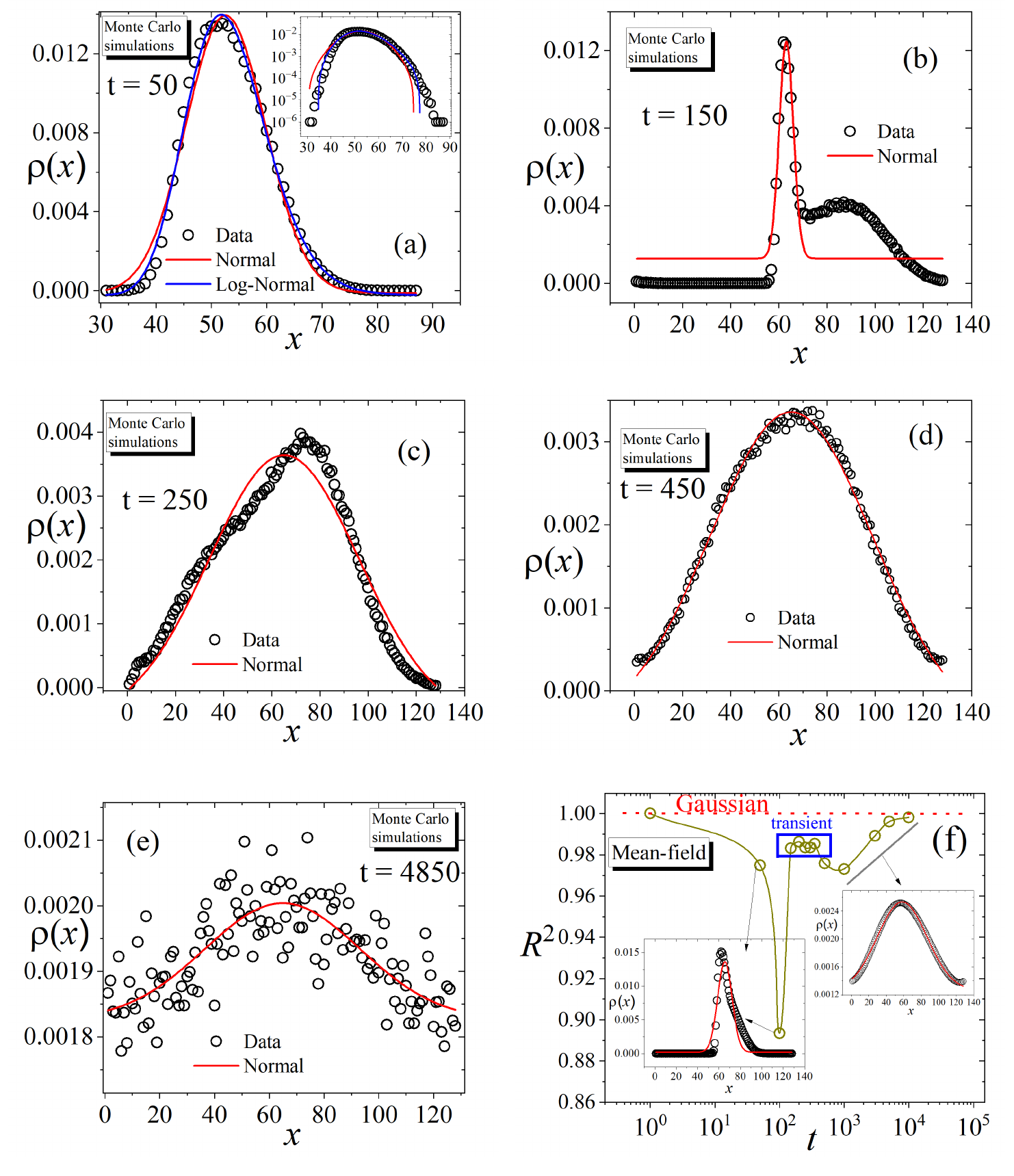}
\end{center}
\caption{ Plots (a)–(e) show the particle distribution at different times obtained via Monte Carlo simulations. At $t = 50$, significant deviations from a normal distribution are observed. As time progresses, the system gradually returns to a Gaussian-like profile, culminating in a noisy steady-state distribution at $t = 400$, as seen in plot (e). Plot (f) summarizes the corresponding results for the mean-field approximation. Here, the quality of the Gaussian fit over time is quantified by the coefficient of determination $R^2$, represented by dark yellow circles. Initially, interactions among particles cause strong deviations, leading to a transient regime. At later times, the system converges again to a Gaussian distribution—less noisy than that of the Monte Carlo steady state. The inset in (a) compares normal and log-normal fits on a semi-logarithmic scale. The insets in (f) illustrate the particle profile at different stages of the evolution.
} \label{fig_gaussian_to_gaussian}
\end{figure}

This transition from one Gaussian state to another is illustrated in Fig.~\ref{fig_gaussian_to_gaussian}. In Figures~\ref{fig_gaussian_to_gaussian}(a)–(e), we show the results from MC simulations, while Fig.~\ref{fig_gaussian_to_gaussian}(f) corresponds to the mean field approximation. All plots refer to the case of $m=4$ species, $n=256$ particles, and interaction parameter $\alpha = 0.249$.

In Figure~\ref{fig_gaussian_to_gaussian}(a), we observe the particle distribution at $t=50$. We observe that the initial Gaussian behavior is disrupted due to particle interactions. The red and blue curves represent normal and log-normal fits, respectively, with the latter providing a more accurate fit at this stage. The inset plot shows the same trend on a semi-log scale. By $t=150$, the distribution becomes strongly non-Gaussian, with a pronounced peak emerging, as shown in Fig.\ref{fig_gaussian_to_gaussian}(b).

At $t=250$, the shape of the distribution becomes nearly triangular (Fig. \ref{fig_gaussian_to_gaussian}(c)), signaling the beginning of a return to Gaussianity. A good Gaussian fit is eventually recovered in Fig.~\ref{fig_gaussian_to_gaussian}(d), and a noisy but stable Gaussian distribution is observed at $t=400$, as shown in Fig.~\ref{fig_gaussian_to_gaussian}(e). This reflects the establishment of a steady state after extensive particles interactions.

Finally, in Fig. \ref{fig_gaussian_to_gaussian}(f), we analyze the mean-field approximation. Unlike the MC simulations, which display the full evolution of the particle profile, we summarize the results using the coefficient of determination $R^2$, which quantifies the quality of a normal distribution fit over time. These values are represented by circles. Initially, we observe a sharp decline in $R^2$, indicating the breakdown of Gaussianity (illustrated in the inset). As $t$ increases, the system goes through a transient (points inside blue rectangle) phase before recovering a Gaussian profile, again shown in the inset. This confirms that the same Gaussian-to-Gaussian behavior occurs in the mean-field scenario, in agreement with the MC simulations presented in Figs.~\ref{fig_gaussian_to_gaussian}(a)–(e).

\section{Conclusions}

In this work, we proposed a mean-field approximation to describe the dynamics of an agent-based model of $m$-species of particles. Specifically, we proposed a recurrence relation that shows the evolution of lattice gas dynamics with an asynchronous updating scheme in a Von Newman neighborhood.

We showed that the MF method shows a good agreement with the particle dynamics implemented through MC simulation using initial conditions and static floor fields carrying polar symmetry in relation to the center of the lattice. We studied the cases of one, two, and four species, which, with the initial conditions, made for a sort of ``controlled'' environment so that we could access more easily the agreement between methods.

The MF approach carried a normalization constant exponent $\beta$, which relied on the parameters $n$ and $m$, because the agreement of methods is strongly influenced by the complexity of the interactions. We showed that when one species is considered, its optimal value, $\beta_c$, has a linear dependence on the total number of particles, when $m=2$ and $m=4$ it shows to be a function of some power of $n$.

We highlighted that our proposed mean-field approach is not bulletproof by showing what specific set of parameters made it show numerical instability due to competing terms inside the recurrence relation.

Despite studying only three cases, our approach is general enough to allow an arbitrary number of particles, even for initial conditions and/or species preferential directions of motion different from those studied in this work. Not only that, our framework proposed by Eq.~\ref{rec_1.1} presents a general feature of application to asynchronous lattice gas models, even reproducing the recurrence relation of other models such as the deterministic two-species dynamics studied in~\cite{Cividini_2013}. 

We are currently exploring higher systems with more than four species, which means that some species are going to present static floor field at angles lower than $90^{\circ}$ with the $x-$ and $y-$direction. As a consequence, we expect to observe $\beta_c$ presenting a more complex dependency with $n$.

Finally, our model reveals a distinctive pattern we term the "Gaussian-to-Gaussian" transition. In this phenomenon, particles initially distributed in a Gaussian configuration undergo significant deformation of this pattern, only to eventually recover a Gaussian distribution after multiple interaction cycles. Both Monte Carlo simulations and mean-field theory reproduce this effect, though they follow different temporal evolution pathways.

\section*{Conflict of Interest Statement}

The authors declare that the research was conducted in the absence of any commercial or financial relationships that could be construed as a potential conflict of interest.

\section*{Author Contributions}

All three authors contributed equally to this work. They jointly conceived and designed the analysis, conducted formal analyses, wrote the paper, elaborated on the algorithms, analyzed the results, and reviewed the manuscript.

\section*{Funding}

We gratefully acknowledge the partial support provided by CNPq: E. V. Stock under grant 153315/2024-5, R. da Silva under grant 304575/2022-4, and S. Gonçalves under grant 314738/2021-5.

\section*{Acknowledgments}

We appreciate the availability of computational resources from the Lovelace cluster at IF-UFRGS.

\section*{Supplemental Data}

All supplemental material can be found at the article webpage.


\bibliographystyle{Frontiers-Vancouver} 
\bibliography{Bibliografia}

\end{document}